\documentclass[twocolumn,showpacs,prl]{revtex4}

\usepackage{graphicx}%
\usepackage{dcolumn}
\usepackage{amsmath}
\usepackage{latexsym}

\begin {document} 
 
\title 
{ 
An optimal network for passenger traffic
} 
\author 
{ 
A.K. Nandi, K. Bhattacharya and S.S. Manna
} 
\affiliation
{ 
Satyendra Nath Bose National Centre for Basic Sciences 
    Block-JD, Sector-III, Salt Lake, Kolkata-700098, India
} 
\begin{abstract}
The optimal solution of an inter-city passenger transport network has been
studied using Zipf's law for the city populations and the Gravity law describing the fluxes of inter-city
passenger traffic. Assuming a fixed value for the cost of transport per person per kilometer we observe 
that while the total traffic cost decreases, the total wiring cost increases with the density of links. 
As a result the total cost to maintain the traffic distribution is optimal at a certain link density which
vanishes on increasing the network size. At a finite link density the network is scale-free. Using this 
model the air-route network of India has been generated and an one-to-one comparison of the nodal degree values
with the real network has been made.
\end{abstract} 
 
\maketitle 
   The identification of certain crucial controlling parameters that ensure the characteristic structure 
of a random network, be it a network that has been created by a natural process or a network that has evolved due to the 
social requirements, has been a focal point of research interest for quite some time
\cite {Barabasi,Dorogov,social,Manna}. For example, 
different algorithms have been proposed to generate the well known scale-free structures of highly
heterogeneous networks which successfully reproduce the statistical features of important networks 
like the Internet \cite {Faloutsos}, World Wide Web \cite {WWW} and airport networks \cite {Barrat} etc.

   On the other hand, not much attention has been paid in reproducing the structural features specific 
to a particular network and to making a one-to-one comparison of the real and the model networks. Intuitively 
it is evident that such a modeling would need information specific to such a network. In this paper we 
argue that for a network of passenger traffic it is possible to construct an optimized model network 
of this kind using only two ingredients, namely the node-wise population distribution as well as a guiding
rule for the passenger traffic flows.

   A transport network should be efficient as well as cost effective. Efficiency is ensured when the
communication between an arbitrary pair of nodes takes only a finite and short duration even when the 
network is very large. This implies that the network must be characterized by `small-world' features.
In addition the network should be robust with respect to 
random failures. If a link is down, the transport process should not be grossly affected. This 
implies that the network must not have a tree structure which is most economic but has extreme
sensitivity to failures. In practice the network should be such that when the flow is not possible 
along a certain path, there must exist alternate paths, even of longer lengths, 
to maintain the flow. Indeed real-world transport 
networks are never like tree graphs. Actually they have multiple loops of many different length scales and therefore 
they are hardly affected by random link or node failures. A prominent example of this is the Internet and its robustness 
to random failures in its structure is quite well known \cite {Albert}. Secondly, the laying cost
of the network is another controlling factor. If every node is connected 
to all other nodes it would be excellent, but that would involve large establishment and maintenance
cost. Planners and administrators of railway networks, city bus transport systems, or even postal networks 
establish and upgrade their networks keeping mainly these two aspects in mind. 

   Recently, optimal networks embedded in Euclidean space have attracted much attention.
Given a spatial distribution of human population the locations of the different facilities so that the mean distance is a minimum was discussed in
\cite {Gastner}. Signatures of topology and patterns are explored in \cite {Gastner1}. A minimal spanning
tree structure of the optimal network was proposed in \cite {Barthelemy}.

   Here we study a model network for the passenger traffic among different cities. 
We ask if, given the populations and locations of all cities in a country, can one predict 
the structure of the network that is optimized with respect to the connection robustness and 
wiring cost? Our study is based on the framework of 
Zipf's law \cite {Zipf,ZipfWiki} of city population distribution and the Gravity law \cite {Tinbergen} 
of social and economic sciences describing the strength of the passenger traffic between a pair of
cities. Finally, we apply this scheme to the Indian air traffic network, which gives good
correspondence with the real network.

\begin{figure}[top] 
\begin{center} 
\includegraphics[width=6.5cm]{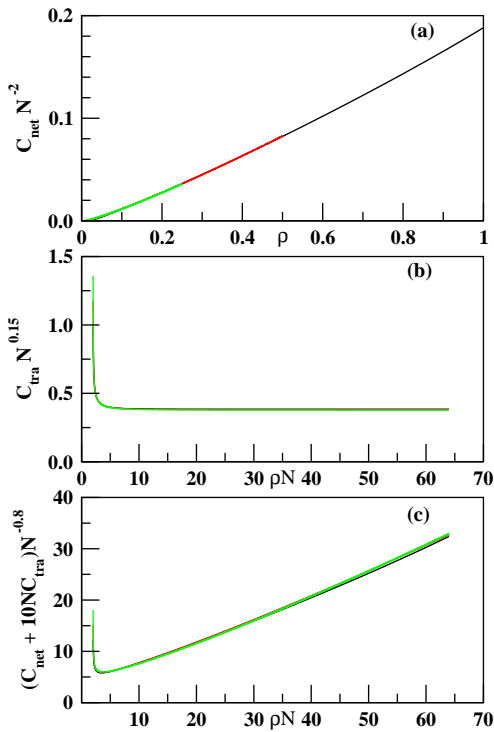} 
\end{center} 
\caption{
Finite-size scaling analysis network cost functions demonstrated by the collapse of the data 
for network sizes $N$ = 128, 256 and 512:
(a) the total wiring cost ${\cal C}_{net}N^{-2}$ scales with the link density $\rho$,
(b) the total travel cost ${\cal C}_{tra}N^{0.15}$ to maintain the traffic scales with $\rho N$ and
(c) the total cost $({\cal C}_{net} + 10NC_{tra})N^{-0.8}$ also scales with $\rho N$.
} 
\end{figure} 
 
   Zipf's law for the frequency of occurrence of English words
has also been applied to the rank-size distribution of city populations \cite {Zipf}. In a country
the maximally populated city is assigned rank 1, the second largest is put in rank 2, etc.
It is known that the population size $p(r)$ varies inversely with rank $r$. This also
implies that the population distribution is a power law \cite {ZipfWiki}.

   The passenger traffic among $N$ different cities and towns in a country
is given by the well-established Gravity model~\cite {Tinbergen}. In its introductory 
form the magnitude of the passenger flow from city $i$ to city $j$ is jointly
proportional to their individual populations $p_i$ and $p_j$ and at the same time is
penalized by an inversely proportional factor which is the square of their distance of 
separation $\ell_{ij}$ as
$F_{ij}  \propto p_ip_j/\ell_{ij}^2$. This equation has been generalized 
to the following asymmetric parametric form ~\cite{Head}:
\begin{equation} 
F_{ij}={p_i^{\alpha}} \left({\frac{p_j^{\beta}}{\ell_{ij}^{\theta}}}/ 
{\sum_{k \ne i} \frac{p_k^{\beta}}{\ell_{ik}^{\theta}}}\right)
\end{equation}
where $\alpha$, $\beta$ and $\theta$ are suitable parameters and $k$ runs over all $N-1$ nodes except $i$.

   While applying these two laws we assume that not only the total 
population of the country is conserved but also the individual city populations
remain constant. More specifically, we assume that in a certain unit of time $F_{ij}$
tourists travel from city $i$ to city $j$ but they eventually return to their
own city $i$ within the same time interval. Of course there are a few who migrate from 
one city to the other and start living there, but their number must be very small 
compared to the tourist traffic, and we ignore this component of migratory flow. Therefore 
neither the city populations nor the inter-city traffic flow changes with time. It seems
that our model should also be quite appropriate for a postal distribution network.

   In a simple model we take a unit square box on the $x-y$ 
plane to represent the country and $N$ points distributed at random positions within 
the box as the locations of different cities. Though the periodic boundary 
condition has no physical meaning in this context we use it along both the transverse 
directions on the box to make the data more well behaved. Given the set of coordinates 
of $N$ points $\{x_i,y_i\}, i=1,N$, all inter-city distances $\ell_{ij}$ are determined.
Cities are then assigned populations $p_i, (i=1,N)$ (in real numbers) by drawing them 
from a power law distribution ${\rm Prob}(p) \sim p^{-\mu}$ with $\mu=1$ as per Zipf's law.
Using $p_{min}=0.001$ and $p_{max}=1$, the city populations are generated using the
relation $p=p_{min}\left(p_{max}/p_{min}\right)^{r_1}$, where $r_1$ is a uniformly 
distributed random fraction and finally normalized such that $\Sigma_ip_i=1$.

\begin{figure}[top] 
\begin{center} 
\includegraphics[width=7.5cm]{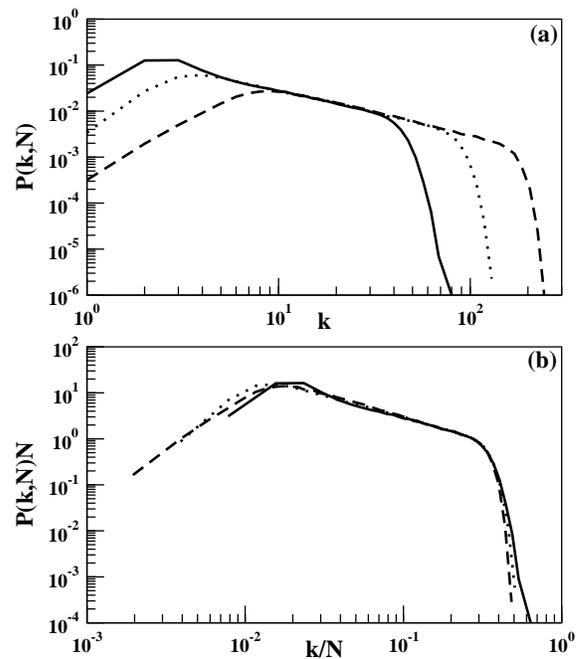} 
\end{center} 
\caption{
The nodal degree distribution $P(k,N)$ vs. $k$ of the optimized network at the link 
density $\rho = 0.1$. (a) The distribution for network sizes $N$ = 128, 256 and 512. 
(b) Finite size scaling of the same distributions gives a value of the exponent $\gamma=1$.
} 
\end{figure} 

   Knowing the values for the city populations $p_i$ and the mutual inter-city distances,
the magnitudes of passenger fluxes $F_{ij}$ and $F_{ji}$s are calculated using Eqn. (1)
and for a certain set of values of $\alpha$, $\beta$ and $\theta$. By definition this flow 
pattern is inherently directed. However, we consider only an undirected traffic flux between 
$i$ and $j$ by considering the net flow $\tilde F_{ij} = F_{ij}+F_{ji}$. Let at some 
arbitrary intermediate stage all $N$ cities be linked by a singly connected network. Since 
nodes are randomly distributed on a continuous plane, there exists one and only one shortest 
path between a pair of nodes. We assume that the entire flow $\tilde F_{ij}$ passes through 
the shortest path on the network connecting the nodes $i$ and $j$ and therefore each link on 
this path is assigned $\tilde F_{ij}$. When this assignment process has been completed for all distinct 
$N(N-1)/2$ node pairs, the net flow through a link measures the net traffic $w$ through that link.
The quantity $w$ is like a weighted betweenness centrality and the net traffic along a link connecting
a pair of nodes $i'$ and $j'$ is $w_{i'j'}=\Sigma \tilde F_{ij}$, where the summation is taken over
the subset of $N(N-1)/2$ node pairs whose shortest paths pass through the link $\{i'j'\}$.
On a graph having loops the shortest paths are found using the well-known Dijkstra algorithm \cite {Dijkstra}.

\begin{figure}[top] 
\begin{center}
\begin {tabular}{cc}
\includegraphics[angle=-90,width=4.5cm]{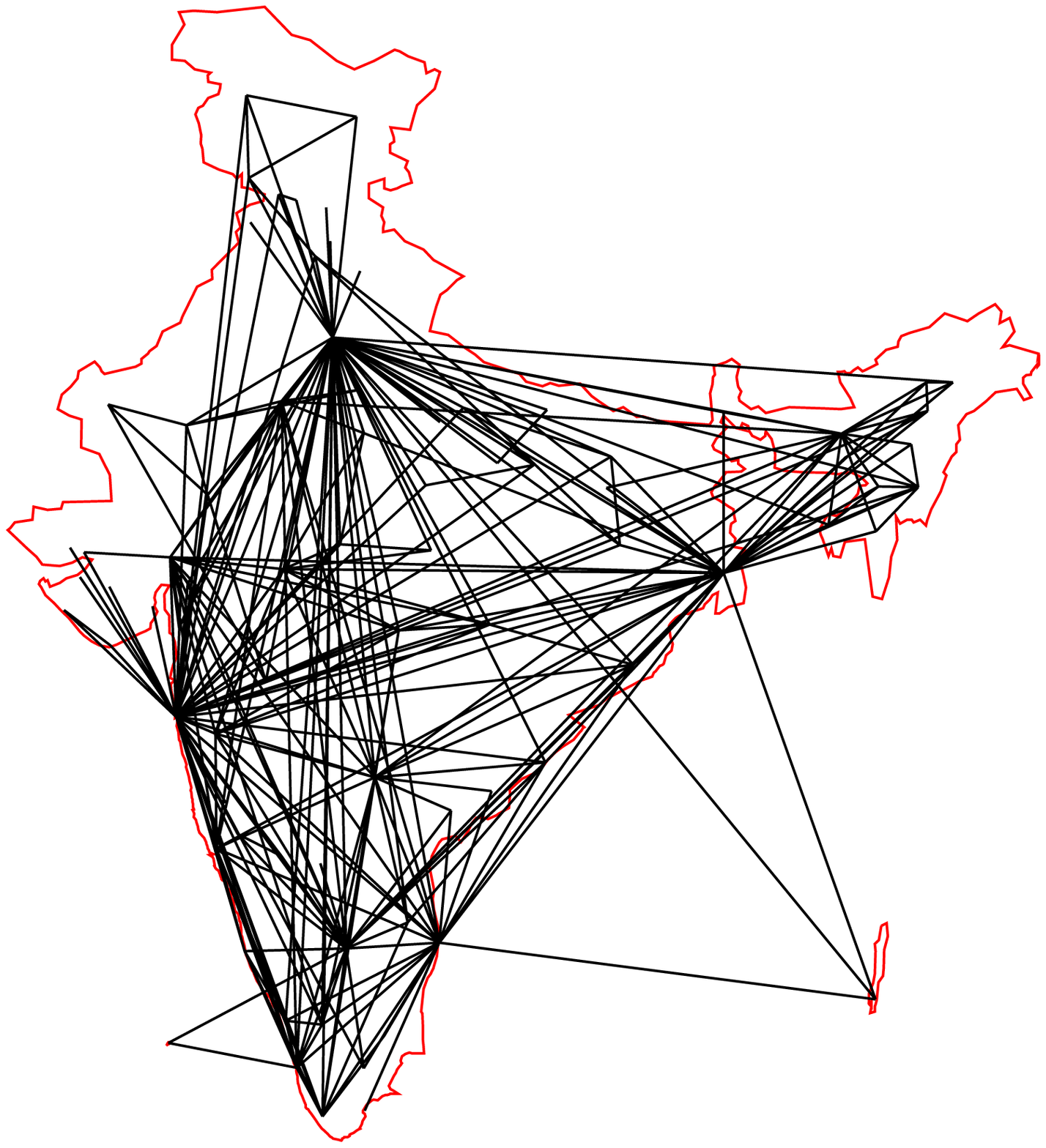} & 
\includegraphics[angle=-90,width=4.5cm]{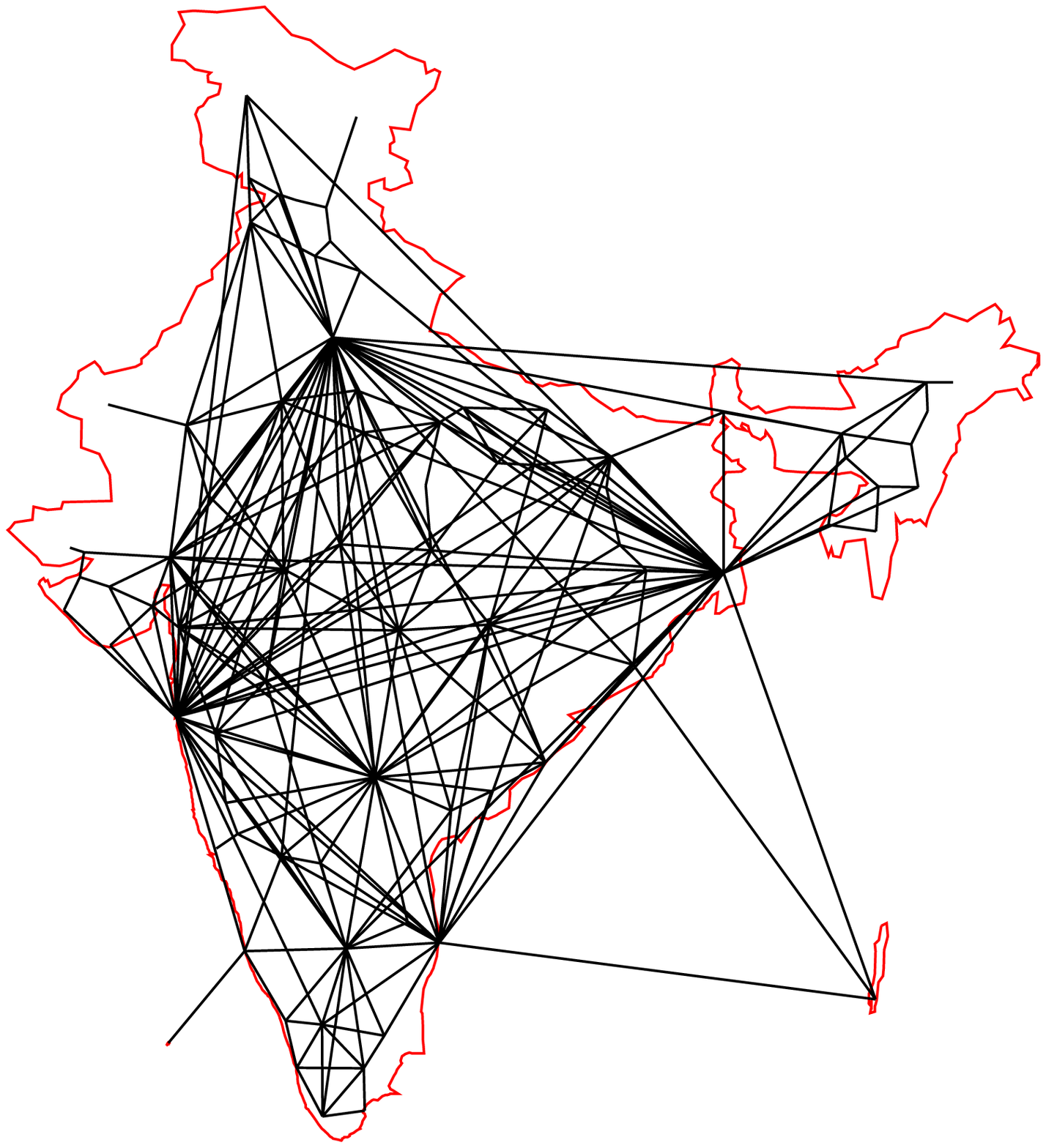} \\
{\bf (a)} & {\bf (b)} \\
\end {tabular}
\end{center} 
\caption{
Air route connection network of 80 civilian airports in India. (a) The 
real air route network of $L$ =265 links connecting different pairs of airports
by all 12 airline companies active in India. (b)
The network obtained from our model using 2001 census data for the city 
populations of the associated Indian cities and using the Gravity law with $\alpha = \beta = 1$ and $\theta=2$.
This network also has 265 links.
} 
\end{figure} 

   The cost function for this traffic distribution has two competing factors. Given a network
there is a cost to maintain the traffic along every link. We assume a fixed value for 
the cost to transport a unit population through unit distance along every link of the network
(e.g., per person per kilometer).
Therefore if $w_{ij}$ is the net flow between the two end nodes $i$ 
and $j$ of a link of length $\ell_{ij}$ then the total cost involved to maintain the entire traffic flow is 
${\cal C}_{tra}=\Sigma_{i\ne j}w_{ij}\ell_{ij}a_{ij}$. The second factor
is the establishment cost to construct the connections, which is,
${\cal C}_{net}=\Sigma_{i\ne j}\ell_{ij}a_{ij}$. Here, the $a_{ij}$s are the elements of the 
adjacency matrix and $a_{ij}=1$ if there exists a link between $i$ and $j$ otherwise it is
0. Therefore the total cost function to maintain the whole traffic distribution of the network is
the sum of these two factors:
\begin {equation}
{\cal C} = {\cal C}_{net} + \lambda {\cal C}_{tra} = \Sigma_{i\ne j} (1 + \lambda w_{ij})\ell_{ij}a_{ij}
\end {equation}
where $\lambda$ is a type of conversion factor that makes an equivalence between the two types of cost.

   The Minimal Spanning Tree (MST) graph covering all $N$ nodes using the Euclidean distances ${\ell_{ij}}$ 
as the link weights has the minimal value of the networking cost ${\cal C}_{net}$. Using Kruskal's 
algorithm \cite {Kruskal} to generate the MST, the whole set of $N(N-1)/2$ links is arranged in 
a sequence of increasing lengths. Links are then dropped one by one from this sequence, starting from the minimal 
length. Links which form loops are rejected. This is checked by the well-known Hoshen-Kopelman algorithm 
in Percolation Theory \cite {HoshenKopelman}. The MST is obtained when one successfully places $N-1$ links.

   To construct the optimal network for passenger traffic we start by constructing the MST as described 
above, connecting all the $N$ nodes and the cost components ${\cal C}_{tra}$, ${\cal C}_{net}$ and ${\cal C}$ 
for this network are calculated. Obviously the optimized network cannot have a tree structure where the 
typical distance between an arbitrary node pair is much larger. Therefore we drop additional links onto the 
graph to minimize ${\cal C}_{tra}$ using the following procedure. First, all node pairs which are not linked 
are sorted out. Our strategy is to pick up the particular unlinked node pair which if connected decreases 
${\cal C}_{tra}$ by the maximal amount. If the length of the shortest path measured on the network between 
a typical pair of unlinked nodes $i$ and $j$ is $d_{ij}$ then the reduction in travel cost when the node pair 
is connected directly by a link of length $\ell_{ij}$ is approximately $\Delta{\cal C}_{tra} = (d_{ij}-\ell_{ij})\tilde F_{ij}$. 
This difference has been calculated for all unlinked node pairs in a similar way. We select the particular node pair for which 
$\Delta {\cal C}_{tra}$ is maximum and link them. After that we recalculate all shortest paths and $w_{ij}$ 
values afresh and repeat the whole process to add the next link. In this way links are added one by one and 
values of both the cost components and the total cost are measured with increasing link density. 

\begin{figure}[top] 
\begin{center} 
\includegraphics[width=7.5cm]{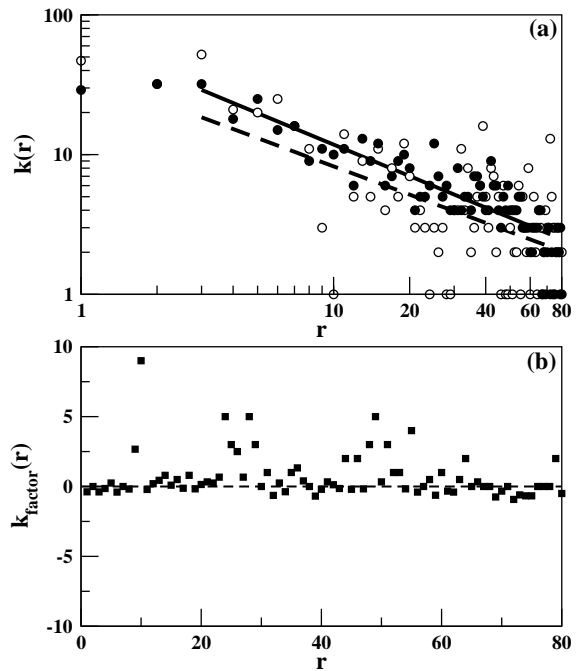}
\end{center} 
\caption{
(a) The degree $k(r)$ of an airport in India has been plotted with its rank $r$ in the population:
real data in open circles and model data in filled circles. Average slopes are shown by the dashed line
and the solid line for the real and model data, respectively.
(b) The one-to-one comparison of the nodal degree values using the $k_{factor}(r)$ between the real
and the model networks. This factor has been plotted with the nodal rank $r$.
}
\end{figure} 

   We first analyze the finite size scaling of the variation of different cost components with the 
link density $\rho(N,L)=2L/\{N(N-1)\}$ using $\alpha=\beta=1$ and $\theta=2$, with $L$ being the number 
of links of the network. In Fig. 1(a) the scaled total wiring cost ${\cal C}_{net}N^{-2}$ increases 
monotonically with the link density $\rho$ for all $N$. In Fig. 1(b) the total traffic cost initially 
decreases very sharply but then almost saturates and ${\cal C}_{tra}N^{0.15}$ scales with $\rho N$. 
Finally, to obtain a scaling of the total cost function we have to use the conversion factor $\lambda = 10N$
and define ${\cal C} = {\cal C}_{net} + 10N{\cal C}_{tra}$, which scales as (Fig. 1(c))
\begin {equation}
{\cal C}N^{-0.8} \sim {\cal F}(\rho N) \nonumber
\end {equation}
where ${\cal F}(x)$ is a scaling function independent of $N$. The total cost so defined first decreases 
and then increases and has a distinct minimum at a specific link density $\rho_c$. We call the network 
at this link density $\rho_c$ the globally optimized traffic solution with respect to the total cost.

    The nodal degree distribution has been calculated on the optimized network at the link density $\rho_c$.
From Fig. 1(c) we see that $\rho_c \to 0$ as $N \to \infty$. Consequently the degree distribution at 
$\rho=\rho_c$ is nearly the same as the Poisson distribution, the degree distribution of the MST. However, 
for finite link density the network becomes even more heterogeneous. This is manifested by the appearance of 
many small degree nodes and few large degree hubs. In Fig. 2(a) we have plotted the degree distribution $P(k,N)$ 
for three different network sizes but at the same link density. The finite size scaling has been shown in 
Fig. 2(b) as
\begin {equation}
P(k,N)N^{\eta} \sim {\cal G}(k/N^{\zeta})
\end {equation}
where $\eta = 1 $ and $\zeta = 1 $ are used to obtain the best data collapse giving the degree distribution
exponent $\gamma = \eta / \zeta = 1$. This clearly shows that at finite link density the resulting network is 
a scale-free network. At a different link density the exponent value is the same and the scaling range increases 
with the density. It is also observed that the model network is a small-world network from the density $\rho_c$ 
and beyond. It has already been reported that the air-route network of India has a scale-free structure \cite {Bagler}. 

   We apply our model to the air-traffic network in India.
There are $N$=80 civilian airports in India for passenger transport. Considering all 12 airline companies \cite {Airlines}
that are active in the air-space of India there are $L$ = 265 links connecting different pairs of airports \cite {Airdata}. 
Every airport is associated with a nearby city or town. We have collected the 2001 census (in India a general 
census is done every ten years) data for the populations of these cities from the website \cite {Census}. We have
checked that indeed the available data for the top 185 populated Indian cities do follow Zipf's law quite closely. 
The mutual distances of separation among these cities are calculated by the arc lengths of the great circles on the 
surface of the Earth joining pairs of cities. The latitudes $\phi$ and longitudes $\psi$ of these cities are 
available at the website \cite {Lat-Long}. Knowing these angles, the inter-city distances are obtained from
\begin {eqnarray}
\ell_{ij} & = & R_E Cos^{-1}([Sin(\phi_i).Sin(\phi_j)] \nonumber \\
          & + & [Cos(\phi_i).Cos(\phi_j).Cos(\psi_i-\psi_j)]) \nonumber
\end {eqnarray}
where $R_E = 3962.6$ miles is the Earth's radius and the angles are measured in radians. As per our scheme we 
start by constructing an MST using $\ell_{ij}$ as the weights. The total of 265 links are then dropped one by one using the
optimized procedure described above until the link density $\approx 0.084$ in the real network is reached. Once the fully 
connected network has been formed we draw the airline networks for both the real as well as model data in 
Fig. 3(a) and 3(b), respectively.
The visual similarity of these two figures is quite striking. However, on a closer look it seems that the links are
more dense for the real network, though this is not true. Though the model network has exactly the same number of links
as the real network, the longer links are more in the real network. This is because we started from an MST, and shorter
links are selected in the very starting configuration. 

   For both networks we measure the nodal degree 
$k_i$ and the rank $r$ of every node. The degree vs. rank plot of the 80 node
networks of India are compared between the real and model networks in Fig. 4(a). The comparison is good, and both 
seem to obey a power law of $k(r) \sim r^{-\nu}$ with $\nu = 0.67$ and 0.75 for the real and model networks. 
Secondly, we study a one-to-one comparison of the real and model
networks. For this purpose we calculate the factor $k_{factor}(r)=(k_{model}(r) - k_{real}(r))/k_{real}(r)$ 
and plot it with the rank $r$ in Fig. 4(b). We observe that $k_{factor}(r)$ remains limited within $\pm 2$ for 80$\%$
of the nodes. We conclude that the correspondence between the real and the model networks is quite good.
To explore the dependence of our results on the parameter values we looked 
at the variation of the quantity $\chi = \Sigma_i (k^i_{model}(r) - k^i_{real}(r))^2$
within the range of $\alpha = \beta = 1 \pm 0.5$ and $\theta = 2 \pm 0.5$,
and observed around 10-15$\%$ variation.

   Finally, we would like to make three comments. (i) There could be a number of ways in which our study 
can be improved. The population in every city is distributed within a wide range of economic 
strengths. Only a fraction of population at the top edge of the distribution generally has access to air 
travel. Therefore perhaps it is better to replace the city population in our study by the number of 
people who are richer than a certain cut-off mark. We thought that an approximate estimate of this number 
would be the number of Income Tax payers. Using these numbers may make the analysis better provided they 
are not proportional to the city populations. Unfortunately we could not get city-wise statistics of 
Income Tax payers. Also, a city airport serves the neighboring towns and suburbs as well. Therefore an 
effective population of the city including its surroundings may be considered. (ii) This point,
which is very interesting, was raised by a referee. In Fig. 4(b) there is a point whose $k_{factor}$ is 9. Such a
high value implies that the model degree is 10 times that of the real degree. We identified the city as Kanpur,
which has a population of $\approx$2.7 million and occupies the 10-th position in the rank distribution. In spite of such
a high population, Kanpur airport is connected only to Delhi and no other airport in the country, leading
to its real degree being unity, but our model predicts it to be 10. A search on the Internet \cite {Kanpur} reveals that indeed
the traffic in Kanpur has increased to a high extent in recent years; there are modernization plans from the government 
and many other airlines are also planning to operate there. Therefore it is expected that the degree of Kanpur 
airport will eventually increase in the near future. This tendency has been correctly predicted in our model.
(iii) The third point has also been raised by another referee.
It may be interesting to explore if our method can be applied to the Indian Railway network as well. Recently it has been
observed that such a network has the small-world property \cite {Railways}. The main difference is that a
large number of stations in a railway network are intermediate stations having two stations on the opposite
sides. Therefore in the framework of our study two cities should be linked only if there is at least one train
that starts at one city and finishes its journey on the other. Perhaps we will take up this study in a future project.

   To summarize, we present evidence that a precise one-to-one reproduction of a network is possible once the
key ingredients controlling the structure of the network are identified. We justify this claim by constructing an optimized 
transport network of inter-city traffic in which traffic is controlled by both the city populations (Zipf's law) 
and by the rule of traffic distribution (Gravity law). The total cost function is determined by two competing factors,
i.e., the cost of maintaining the traffic and the establishment cost of the network. This procedure has been 
applied to the airport network of India having 80 civilian airports, and a node-to-node comparison of the nodal degree 
values between the real network and the model network is made. The correspondence is found to be very good.
 
E-mail: manna@bose.res.in

\end {document}